\documentclass[aps,prd,twocolumn,nofootinbib,superscriptaddress,showpacs,amsmath]{revtex4-1}
\usepackage{graphicx,epsf,amssymb,url,hyperref}
\usepackage{mathrsfs}
\usepackage[]{latexsym}
\usepackage[USenglish]{babel}
\usepackage{epsfig}
\usepackage[utf8]{inputenc}

\newcommand{\ve}{\delta}
\newcommand{\be}{\begin{equation}}
\newcommand{\ee}{\end{equation}}
\newcommand{\pro}[2]{\mbox{$\left(  #1 \,,\, #2\,\right) $}}

\newcommand{\PB}[2]{\mbox{$\lbrace\,#1\,,#2\,\rbrace$}}
\newcommand{\QC}[2]{\mbox{$[ \,#1\,,#2\,] $}}

\renewcommand{\th}{\hat \theta}

\newcommand{\wfun}{\Psi[\rho,\lambda]}
\def\comment#1{}

\newtheorem{theorem}{Theorem}

\newtheorem{preposition}{Preposition}
\begin{document}

\title{Quantum no-singularity theorem from geometric flows}
\author{Salwa~Alsaleh}
\email{salwams@ksu.edu.sa}
\affiliation{ Department of Physics and Astronomy, King Saud University, Riyadh 11451, Saudi Arabia}
%
\author{Lina Alasfar}
\email{lina.alasfar3@outlook.fr}
\affiliation{ Universit\'{e} Clermont Auvergne, 4, Avenue Blaise Pascal 	63178 Aubi\`{e}re Cedex, France  }
%
\author{ Mir Faizal}
\email{mirfaizalmir@gmail.com}
\affiliation{Irving K. Barber School of Arts and Sciences, University of British Columbia – Okanagan, Kelowna, British Columbia V1V 1V7, Canada}
\affiliation{Department of Physics and Astronomy, University of Lethbridge, Lethbridge, Alberta T1K 3M4, Canada}
%
\author{Ahmed~Farag~ Ali}
\email{Ahmed.faragali@nias.knaw.nl}
\affiliation{Netherlands Institute for Advanced Study, Korte Spinhuissteeg 3, 1012 CG Amsterdam, Netherlands}
\affiliation{Department of Physics, Faculty of Science, Benha University, Benha, 13518, Egypt}
%

\date{\today}
            
\begin{abstract}
 In this paper, we  analyze the classical geometric flow as a 
 dynamical system. We obtain an action for this system, such that its equation of 
 motion is the Raychaudhuri equation. This action will be used to quantize this system. 
As the Raychaudhuri equation is the basis for deriving the singularity theorems, 
we will be able to understand the effects such a quantization 
will have on the classical  singularity theorems.
Thus, quantizing the geometric flow, we can demonstrate that a 
quantum space-time is complete (non-singular). 
This is because the existence of a conjugate point is a necessary condition 
for the occurrence of singularities, and we will be able to demonstrate that 
such conjugate points cannot occur due to such quantum effects. 
\end{abstract}
\pacs{02.30.Tb, 04.20.Cv, 04.20.Dw,04.60.-m, 04.60.Ds, 03.65.-w}
\maketitle
\section{Introduction}
%
%
\par \noindent Even though general relativity~(GR) 
is one of the most well tested theories, it predicts its own breakdown due to the occurrence of 
singularities. Furthermore, the Penrose-Hawking singularity theorems demonstrate that the 
occurrence of these singularities is an intrinsic property built  
into the structure of classical 
GR, and not a mathematical artifact~\cite{Hawking:1969sw,Penrose:1964wq}.
At the singularities the space-time is no longer a smooth manifold, and the laws of physics 
cannot be meaningful when studied at these points.
Thus, the removal of such singularities is very important physically, 
and it is generally argued that  singularities should be removed due to
quantum gravitational effects.
\par \noindent It has been argued that the problem would be resolved due to
string-theoretical effects  like the  propagation of strings across 
singularities~\cite{2}, colliding Branes in Heterotic M-theory~\cite{4},
string theoretical effect in a Kasner background~\cite{8}, string gas cosmology~\cite{9}, 
and bouncing branes with negative-tension~\cite{10}.  
All these cosmological models are motivated from  string theory, however, the
string theory also predicts the existence of higher dimensions and supersymmetry, 
both of which have not been experimentally observed~\cite{s1, s2}.  
Similarly, it has been argued that loop quantum gravity~(LQG) can also lead to resolution of 
singularities~\cite{l1, l2, l4, l5,l6}, but      it  has  not  been
possible to recover the Einstein equation in LQG~\cite{l0}. 
Hence, the physical validity of these models that are based on string theory and LQG can 
 be questioned. Even though the singularities have also been removed using 
 other phenomenological approaches to quantum gravity, such as  the existence of a
 minimum measurable length scale~\cite{mini}, a modified Wheeler-DeWitt equation~\cite{mi}, 
 quantum gravity condensates~\cite{conde}, and information-theoretic networks~\cite{newt},
 all of these approaches have problems associated with them. Furthermore, all the work on removal 
 of singularities by quantum effects has been done using 
different proposals, and all of which depend on the specifics of a particular
approach to quantum gravity. However, rarely, 
there were attempts to show quantum completeness of space-time without 
referring to a particular model of gravity~\cite{Hofmann:2015xga}.
\par \noindent A quantum mechanical counterpart of the singularity theorems  
would be needed to rigorously prove the absence of singularities and 
completion of quantum space-time, in a model-independent approach. 
It may be noted that the classical Penrose-Hawking singularity theorems were derived using the 
the Raychaudhuri equation~(RE)~\cite{Raychaudhuri:1953yv}, thus it is expected that 
a quantum mechanical generalization of the RE could be used to analyze 
the quantum mechanical version of the classical Penrose-Hawking singularity theorems. 
In fact, semi-classical corrections to the RE from Bohmian trajectories has been 
constructed~\cite{Das:2013oda}, and it has been argued that such an equation can
resolve singularities in cosmology~\cite{Ali:2014qla}, 
and in black holes~\cite{Ali:2015tva}. These trajectories get corrected because 
of the quantum   corrections to the flow of geometries. 
Furthermore, to derive the quantum 
version of the classical Penrose-Hawking singularity theorems, we
would need to understand the full quantum mechanical behavior of the geometric flows.
\par \noindent In this paper,
we shall study the dynamics of these geometric flows. We will construct a classical
action for these geometric flows, such that the classical equation of motion of this 
action would be the Raychaudhuri equation. Then we will quantize this system using the standard 
canonical quantization techniques.  We will show that it is  possible to
study conjugate points and  singularities in quantum space-time
using these quantum geometric flows.
We  will   demonstrate that quantum gravitational effects will indeed
remove the singularities, and we will obtain quantum   no-singularity theorems. 
The classical Penrose-Hawking singularity theorems will be  recovered in the Ehrenfest limit of 
these quantum quantum   no-singularity  theorems. 
\section{Geometric Flows\label{GF}}
\par
\begin{figure}
\centering
\includegraphics[width=0.65\linewidth]{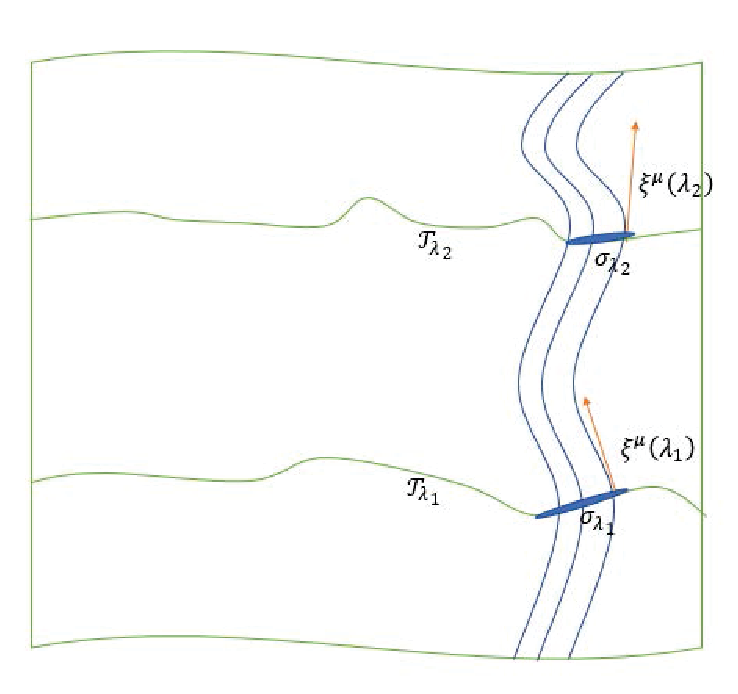}
\label{df}
\caption{The dynamic foliation of the space-time~$\mathscr M$ by the flow 
of geodesic congruence. The cross-sectional hypersurface~$\sigma_\lambda$ represents
the geometric flow with its volume ~$\rho$ being its only dynamic degree of freedom.  
We define $\mathscr T$ as the equivalent classes of~$n$-manifolds that have a
metric~$h_{\alpha \beta}$, whose determinant is equal to~$\rho$ at~$\sigma$.}
\end{figure}

 \noindent We start by studying a congruence
of test particle moving on an~$n+1$ dimensional space-time~$\mathcal M$. So, we  
can use their proper time~$\lambda$ as a dynamical foliation parameter, such that we foliate 
the space-time into the topology~$ \mathscr T \times \mathbb{R}$, 
see Figure~\ref{df}. 
Observe that~$\mathscr T$ is a Riemannian manifold with a metric~$ h_{\alpha \beta}$,
that projects any vector field~$ X^\mu$ onto~$ \mathscr T$. We 
also define~$ \sigma_\lambda\in \mathscr T$ a hypersurface in the transverse manifold
that is defined by congruence intersection of~$\mathscr T$ at a given time~$ \lambda$.  
The volume of that hypersurface is given by
\be
\textbf{Vol.} = \int _{\sigma_\lambda} \sqrt{\det h} d^nx. 
\label{volume}
\ee
\par \noindent We choose the velocity field  of the test particles in the congruence~$ \xi^\mu$ to 
be normal to the~$n$-dimensional transverse manifold~$\mathscr T$. Such that the vorticity would
vanish~\cite{ref1}.
We consider the cross-sectional hypersurface~$ \sigma_\lambda $ as a dynamical system, with its volume as its dynamical degree of freedom. For convenience, we define the dynamical degree of freedom~$ \rho(\lambda) = 2 \int\ \sqrt{\det(h_{\alpha\beta})}$.
Hence, the dynamical variable is not a field variable, but still represents an extended object.\\
We define the dynamical evolution of the transverse
metric~$ h_{\alpha \beta}$ by the equation~\cite{ref1},
\begin{align}
\partial_\lambda h_{\alpha \beta} &= \theta_{\alpha \beta} \nonumber\\
&= 2 \sigma_{\alpha \beta } + \frac{2}{n} h_{\alpha \beta} \theta . \label{geometricflow}
\end{align}
With~$ \sigma_{\alpha \beta}$ being the traceless shear tensor, and~$\theta $ is the expansion parameter. This equation is a form of a geometric flow equation for the equivalent class of manifolds~$ \mathscr T$, or more accurately for~$ \sigma_\lambda$. 
We may take the trace  of~\eqref{geometricflow} to get,
\be
h^{\alpha \beta} \, \partial_\lambda h_{\alpha \beta} = \frac{2}{n}\, \theta.
\ee
Then, we multiply by~$\sqrt{\det h}$, obtaining:
\be
\sqrt{\det h} h^{\alpha \beta}  \partial_\lambda h_{\alpha \beta} = \frac{2}{n} \sqrt{\det h}\, \theta.
\ee
Using the identity~$\ve (\det h) = h h^{\alpha \beta} \ve h_{\alpha \beta}$ we identify the LHS of above equation with~$ \dot \rho$. Thus we have,
\be
\dot \rho = \frac{2}{n} \rho\, \theta.
\label{rdot}
\ee
%
 %
 %
 %
 \par \noindent  We now make the following ansatz about the action for the geometric flow
\be
S[ \rho, \dot \rho] =\frac{1}{2 \kappa} \int d\lambda \frac{n}{4}\,\frac{1}{\rho}\, \dot \rho^2 \, - \mathcal R\, \rho -  V_\sigma[\rho]  ,
\label{action}
\ee
with~$\mathcal{R}$, being the Raychaudhuri scalar~$\mathcal{R}:= R_{\mu \nu} \xi^\mu\xi^\nu$, $ \kappa = 8 \pi G_N$, and~$V_\sigma[rho]$ is the shear potential, satisfying:
\be
\frac{\ve V_\sigma[\rho]}{\ve \rho} = 2\sigma^2
\ee
We may identify the canonical conjugate momentum to~$\rho$ using the Lagrangian,
\begin{align}
\Pi &= \frac{\delta L}{\delta \dot \rho}  =\frac{n}{2}\, \rho^{-1} \dot \rho, \nonumber \\
&= \frac{n}{2}\, \rho^{-1} (\frac{2}{n} \rho\, \theta) = \frac{1}{2 \kappa}\theta.
\end{align}
Thus, as expected, the expansion parameter is the conjugate momentum to
the dynamical degree of freedom. We proceed further by computing the variation~$
\delta L/\delta \rho$
\begin{align}
2 \kappa\frac{\delta L}{\delta \rho} =&-\frac{n}{4}\rho^{-2} \dot \rho^2 - 2 \sigma^2 - \mathcal R \nonumber \\
=& \frac{n}{4} \left[ \rho^{-2} \left(\frac{4}{n^2} \rho^2 \theta^2 \right) \right]- 2 \sigma^2 - \mathcal R \nonumber \\
=& -\frac{1}{n}\, \theta^2 - 2 \sigma^2 - \mathcal R 
\end{align}
Now, we write the Euler-Lagrange equations in the~$\rho$ configuration space,
\be
\frac{d}{d \lambda}\, \theta = 2 \kappa\frac{\delta L}{\delta \rho}
\ee
Thus, we obtain  the Raychaudhuri equation,
\be
\dot \theta = -\frac{1}{n}\, \theta^2 - 2 \sigma^2 - \mathcal R
\ee
This is the expected result, since the dynamics of the geometric flow should generate the dynamics of the congruence `above' it. Indicating that the action~\eqref{action} is the correct ansatz about the dynamical description of the geometric flows. 
\par \noindent We can moreover write the effective Hamiltonian for geometric flows by preforming a Legendre transformation on the Lagrangian~$L$, The effective Hamiltonian can be written as:
\be
H= \frac{1}{2 \kappa} \left( \left(  \frac{1}{n} \rho \theta ^2 \right)  \mathcal R \rho + V_\sigma\right) 
\ee
Raychaudhuri equation can also be recovered from the Poisson brackets 
\be
\dot{\theta} = - \PB{\theta}{H} =  -\frac{1}{n}\, \theta^2 - 2 \sigma^2 - \mathcal R,
\label{hamltoeq2}
\ee
We shall use the canonical formalism for the geometric flows dynamics in order to canonically quantize it,  as in the next section.
%
%
\section{Canonical Quantization}
%
%
\par \noindent We define the operators~$\hat \rho$ and~$\th$ acting on the Hilbert
space of geometric flows~$\mathscr H$, and the  geometric flow state~$\Psi$. 
We may use the~$\rho$-representation for these operators, such that the Hilbert space 
is identified to be~$ \mathscr H := L^2( \mathbb{R}^+;d\mu[\rho])$,
since the configuration space for~$\rho$ consists only of non-negative values~\footnote{
The determinant of a Remaninan manifold is non-negative. Hence, there is no need to
assume any boundary conditions for~$\rho<0$.}.
This Hilbert space is equipped with the measure~$ d\mu[\rho]:= \rho\, d\rho$.  
The states become wave functions of~$\rho$ and time~$\wfun$. 
In fact, they are wave functionals of the coordinates on $ \sigma_\lambda$
as they are defined by,
\be
\wfun:= \int _{\sigma_\lambda} \psi (\rho(x^\alpha,\lambda) )\sqrt{h} d^nx. 
\ee
%
\par \noindent The pair~$\hat \rho$ and~$\th$ are self-adjoint operators,  that satisfy the canonical commutation relations (CCR), 
\be
\QC{\hat \rho}{\th}=i 2 \kappa \hbar \hat I
\ee
In the~$\rho$-representation, they are identified with~\cite{Reed:1980fa,DeWitt}
\begin{subequations}
\begin{align}
 \hat \rho & = \rho : \mathscr I ( \mathbb{R}^+) \to\mathscr I ( \mathbb{R}^+) ,
  \\ \th & = -i2\kappa \hbar \, \frac{\ve}{\ve \rho} :  \mathscr I ( \mathbb{R}^+) \to\mathscr I ( \mathbb{R}^+).
\end{align}
\end{subequations}
Here,~$\mathscr I (\mathbb{R}^+)$ is a subset of~$L^2( \mathbb{R}^+)$ and~$\wfun$ belongs to it. Now, for this wavefunction to be valid for solving Schr\"{o}dinger's equation~$\mathscr I ( \mathbb{R}^+)$, it should be at least~$ C^2(\mathbb C)$. 
We therefore have a well-defined Hilbert space and operators as endomorphisms acting on it.  We can write the effective Hamiltonian operator,
\be
\hat H= \frac{1}{2 \kappa} \left( \left(  \frac{1}{n} \frac{\ve}{\ve \rho} \rho \frac{\ve}{\ve \rho}  \right)  \mathcal R \rho + V_\sigma\right) 
\ee
and the  Schr\"{o}dinger-like equation,
\begin{equation}
\left(  \frac{-4\hbar^2 \kappa}{n} \frac{\ve}{\ve \rho} \rho \frac{\ve}{\ve \rho} \right)  \Psi + V \Psi = -i\hbar \frac{\partial}{\partial \lambda } \Psi
\label{SRD}
\end{equation}
With
\begin{equation}
V= \frac{1}{2 \kappa} \left(  \mathcal R \rho + V_\sigma\right) 
\end{equation}
\par \noindent This equation is very similar equation to Wheeler-DeWitt's~\cite{DeWitt:1967yk},
in terms that the wave function~$\wfun$ is a wave functional of an extended object of 
a background geometry. Nevertheless, as this equation is for the quantum flow of geometries,
that is a subsystem of the universe not  the universe as a whole. Here, we are studying the space-time locally, as an ensemble of `atoms' of geometry.
In addition, this equation contains a real effective Hamiltonian, 
not a Hamiltonian constraint. Hence, it contains an evolution parameter $\lambda$, 
which acts as a real time for this system. Thus, this equation does not have the problem of time 
that is associated with the usual Wheeler-Dewitt equation. Furthermore, here the foliation 
of space-time is based on the flows of geodesics to a longitudinal and transverse directions, 
relative to the congruence. Unlike the standard ADM-foliation that foliates the 
whole space-time, arbitrarily~\cite{Arnowitt:1962hi}.
%
%
 \section{Conjugate Points and Singularities}
 %
 %
\par \noindent  Since the RE is the equation of motion of the geometric flow, 
we shall discuss the singularity theorems using the language of geometric flows. 
Therefore, we will  discuss  singularities in the dynamical language of geometric flows.
Using the  dynamical foliation adopted at the beginning of section~\ref{GF}, 
we have Global hyperbolicity for~$\mathcal{M}$~\cite{ref1} implied from using the
dynamical foliation. Thus, the strong causal conditions hold for~$ \mathcal{M}$. 
Now, we can recall the following preposition~\cite{Hawking:1969sw,Penrose:1964wq}, 
%
  \begin{preposition}
  If the strong causal conditions hold on~$\mathcal{M}$, it implies that the strong energy conditions hold as well, 
  \[ \mathcal{R} >0 \]
  \end{preposition}
%

Which leads us to the focusing theorem~\cite{Borde:1987qr}
%
\begin{theorem}
If the strong energy conditions hold on~$\mathcal{M}$, then a congruence of time-like geodesics will encounter a conjugate point/caustic at a time not greater than~$ \frac{n}{\theta_0}$. For a given initial value of the expansion parameter~$\theta_0 <0$. That is, if congruence enters an isolated horizon.
\end{theorem}
%
The implication of this theorem to the geometric flows is that,  if the initial canonical momentum was less than zero, the geometric flow will evolve in time such that the cross sectional volume decreases quickly and reaches zero~$\rho \to 0$, when~$\theta \to -\infty$, and~$ \lambda = n/\theta_0$. 
\par \noindent The existence of a conjugate point does not imply that the geometric flow hits a singularity, because its volume might expand again after sometime, if~$\dot \theta$ is defined at that point. However, the existence of a conjugate point is an essential step for showing the existence of a singularity~\cite{ref1}.
\par \noindent We shall not discuss the classical singularity theorems in terms of geometric flows any further, as they are kept for later extended study. However, we shall demonstrate the absence of  conjugate points, due to quantum effects. As the existence of  conjugate points is a necessary condition for the occurrence of singularities, we will be able to demonstrate that singularities will not form in quantum space-time. 
%
%
\section{Completeness of Quantum Space-Time}
%
%
\par \noindent Now, we turn to  the existence of conjugate points in the context of quantum geometric flows. The focusing theorem stated above will only hold in the Ehrenfest limit for the operator~$\th$.  We need to investigate this theorem  for the operator~$\th$ without referring to the classical argument. 
\par \noindent The CCR relation between~$\hat \rho$ and~$\th$ leads to the uncertainty relation,
\be
 \Delta \rho \Delta \theta \geq \frac{\hbar}{2}.
 \label{uncertainty}
\ee
Indicating that quantum geometric flows do not tend to focus forming conjugate points, due to the uncertainty in measuring the volume, without causing them to expand rapidly, indicating that gravity at the small scale~$\sim \ell_p$ possess a quantum repulsive force.
\par \noindent In order to show that conjugate points do not form for quantum geometric flows, we need to show that the spectrum of the operator $ \varsigma (\th) $ is bounded below. This can be done by showing that the operator itself is bounded from below~\cite{kato1995}, i.e.
\begin{equation}
\left| \pro{\Psi}{\th\,\Psi} \right|  \geq c,
\end{equation}
for some constant $c$.\\
Such that the expansion will not blow down to minus infinity, as opposed to  the classical case. 
We know that the spectrum of~$\hat \rho$ is bounded below since it takes only non-negative values~$ \varsigma( \hat \rho ) \in \mathbb{R}^+$. Thus, the Hilbert space is, as mentioned earlier, the space~$ \mathscr H :=L^2(\mathbb R^+, d\mu)$. We define the inner product on the Hilbert space~$ \pro{\cdot}{\cdot} : \mathscr H \times \mathscr H \to \mathbb R$ by~\cite{Reed:1980fa}
\be
\pro{\Phi}{\Psi} :=  \int_{0}^{\infty} \Phi[\rho;\lambda]^* \wfun\;\, d\mu[\rho]\,
\ee
and the norm from this inner product:
\be
\| \Psi \| = \left| \pro{\Phi}{\Psi}\right| 
\ee
Therefore, we define the norm of the operator~$\th$ :
\be
\|\th \Psi\| =  \left|\,i\, \hbar \int_0^\infty d\mu[\rho]\,\Psi^*_{,\rho}\, \Psi_{,\rho} \right| 
\ee
with the comma denoting the differentiation.
However, the norm of~$\Psi$  satisfies the inequality~\cite{erdogan}
\be
\|\Psi\|^2 = \left| \int_0^\infty d\mu[\rho] \, \Psi^*\Psi \right| \leq \int_0^\infty  d\mu[\rho] \left|  \Psi \right|^2  \leq \;\sup_{\rho \in \mathbb R^+} | \Psi|^2
\ee
We can use the uniform norm to obtain:
\be
\|\hat \theta \Psi\| \geq \| \Psi\| \Rightarrow \|\hat \theta \Psi\| \geq \sup_{\rho \in \mathbb R^+} | \Psi|
\ee
 That proves the operator~$\th$, is bounded below, hence its spectrum~$ \varsigma (\th)  > - \infty$, proving that  conjugate points do not exist in space-time with quantum geometric flows.
\par \noindent This can be understood easily if we recalled that the probability measure density~$ \rho |\Psi|^2$ should vanish at the end points~$ 0$ and~$ +\infty$ in order to satisfy the Born conditions. Hence, the wave function of the geometric flow will give vanishing probability at the conjugate point~$ \rho=0$, implying that singularities do not form in quantum space-time.
The previous analysis was made for congruence of time-like geodesics, but it can be repeated for the null geodesics with letting~$n \to n-1$, and considering the optical RE~\cite{ref1}. 
%
%
\section{No-Singularity Theorem}
%
%
\par
\begin{figure}[h!]
	\centering
	\includegraphics[width=0.7\linewidth]{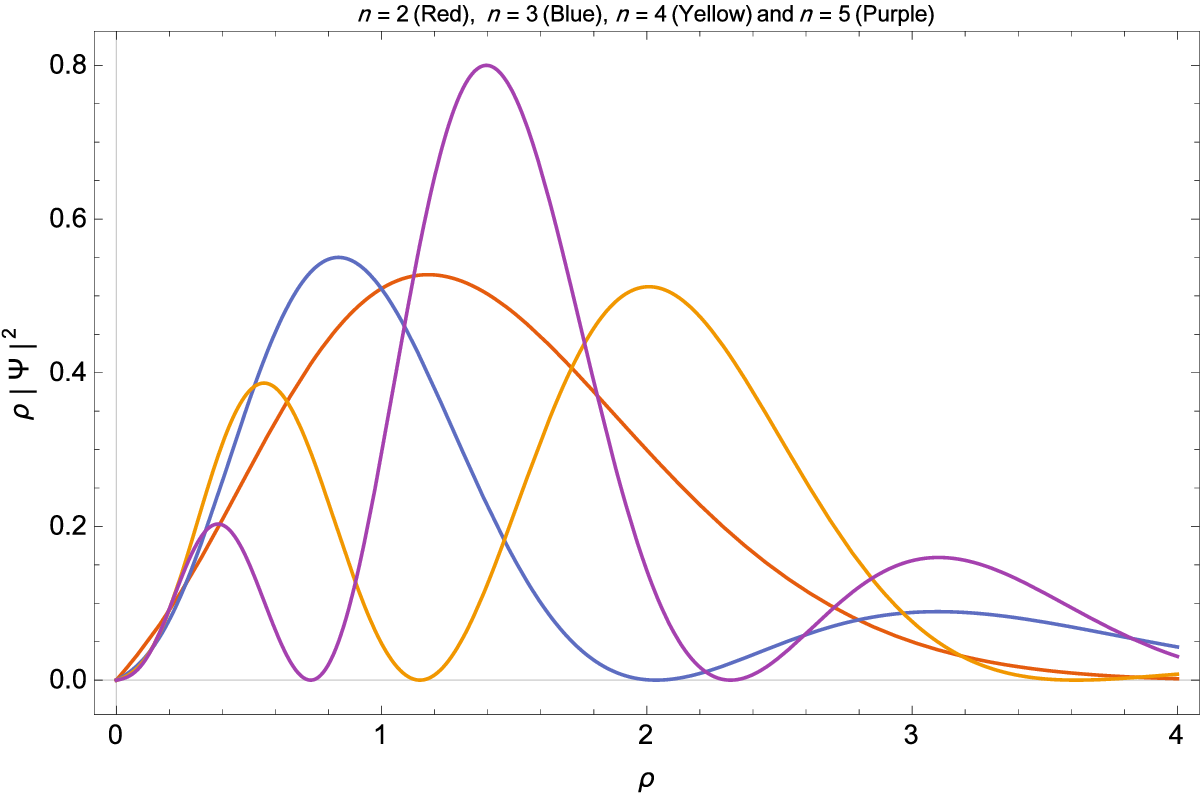}
	\label{cp}
	\caption{ Plots of the probability density functions~$\rho|\Psi|^2$ vs~$\rho$ 
	for different space-time dimensions, obtained by solving~\eqref{SRD} 
	with~$\mathcal{R}=cosnt.$. The plot indicates that the 
	probability measure density function rapidly decreases as~$ \rho\to0$
	and vanishes identically at the conjugate point.}
\end{figure}

 \noindent It is possible to state a `no-singularity' theorem 
for a quantum space-time, from the argument in the previous section.
\begin{theorem}
There is a zero probability for the quantum space-time to have incomplete geodesics due to  singularities.
Hence, the quantum space-time is complete
\end{theorem}
\paragraph{Proof} Although the condition for the operator~$\th$ to be bounded-below 
is a sufficient to prove the above theorem. We can use the Schr\"{o}dinger-like
equation~\eqref{SRD}  to have a more detailed proof.\\ We observe that~\eqref{SRD} 
has an effective mass term
\begin{eqnarray}
m_{eff}= \frac{n}{4} \frac{1}{2\kappa} \rho^{-1}
\end{eqnarray}
this mass diverges at conjugate points
This will affect the behavior of the wavefunction near~$ \rho \to 0$, 
acting like a boundary condition, from a diverging effective potential. 
Therefore, there will be a vanishing probability to measure the geometry 
forming a conjugate point, and conversely a singularity, see Figure~\ref{cp}.
Due to  the effective mass diverging at singularities, we may conjecture that 
the curvature~$ \mathcal R$ is regular even close to the singularity. 
Even if this is not the case and~$ \mathcal R \to \infty
$\footnote{We have not attempted to quantize the curvature directly here, 
this is left for future work.} at the `classical singularity'. 
This could be translated as an infinite potential barrier at~$\rho$, 
forming yet another boundary condition for the wavefunction, such that~$\Psi[0] =0$,
and the above theorem still remains valid. 
%
%
\section{Conclusion}
%
%
\par \noindent In this paper, we have introduced a new dynamical system for the space-time,
the geometric flow associated with the motion of a congruence of time-like or null geodesics
along an~$n+1$ dimensional globally hyperbolic space-time. We have identified the dynamical
degree of freedom for such system being the cross-sectional volume of the congruence at a
given time~$ \rho(\lambda)$. Then, we wrote the action for such degree of freedom.
Variation of the action with respect to~$ \rho$ yielded the Raychaudhuri equation, as
the dynamics of the geometric flows should coincide with the dynamics of the congruence 
described by Raychaudhuri equation.  Then, we identified the conjugate momentum
to~$ \rho$ being the expansion parameter~$ \theta$, and wrote the effective Hamiltonian 
from the Legendre transformation of the Lagrangian. The system was canonically quantized using the
standard techniques.
\par \noindent We were be able to show  from studying the quantum geometric flows that
quantum mechanically the space-time is free from singularities, and these singularities
occur only in the Ehrenfest limit. Thus, we were able to obtain quantum no-singularity
theorem, such that the classical Penrose-Hawking singularity theorems where 
the Ehrenfest limits of such no-singularity theorems. The proof of completeness of 
quantum space-time is made by  showing that the expansion operator~$\th$ 
is bounded from below. Moreover, analysis of the wavefunction~$\wfun$ as a solution
to~\eqref{SRD} with a diverging effective mass at~$\rho=0$ explicitly shows a
null probability for the congruence to form a conjugate point, and thereby 
showing that quantum space-time is complete. Although there is a real singularity
in the space-time manifold~$\mathscr M$, there is a different behavior in
the~$\rho$ configuration space.
Due to the diverging effective mass, 
there is an essential singularity at $\rho=0$.
However, the solution to Schr\"{o}dinger-like equation~\eqref{SRD} has a
non-essential~(removable) singularity at~$ \rho =0$~\cite{arfken}.
Therefore, we expect that analytic extension to the dynamical variable would indicated 
the presence of pre-singular geometry, without loosing regularity. 
The absence of singularity means the absence of inconsistency 
in the laws of nature describing our universe, that shows a particular
importance in studying black holes~\cite{Hofmann:2016vix} and cosmology. 
 %
\section*{Acknowledgments}
\par \noindent We would like to thank Saurya Das, Roberto Casadio and Debasis Mondal 
for their useful discussions and comments. 
This research project was supported by a grant from the ``Research Center of the 
Female Scientific and Medical Colleges'', Deanship of Scientific Research, King Saud University.
%
%


\end{document}